\documentclass[journal=aamick,manuscript=article,email=true,layout=traditional,amsmath,amssymb,linenumbers]{achemso}

\usepackage[version=3]{mhchem} 
\usepackage[utf8]{inputenc}
\usepackage{bm}
\usepackage{graphicx}
\usepackage{hyperref}
\usepackage{color}
\usepackage{nicefrac}
\usepackage{booktabs}
\usepackage{xr}
\usepackage{cleveref}
\usepackage{xcolor}

\setkeys{acs}{maxauthors=0}

\providecommand{\keywords}[1]
{
  \textbf{\textit{Keywords---}} #1
}

\author{Francesco Delodovici}
\affiliation{Université Paris-Saclay, CentraleSupélec, CNRS, Laboratoire SPMS, 91190, Gif-sur-Yvette, France}
\email{francesco.delodovici@centralesupelec.fr}

\author{Charles Paillard}
\affiliation{Smart Ferroic Materials center, Institute of Nanosciences \& Engineering and Department of Physics, University of Arkansas, Fayetteville, Arkansas 72701, USA}
\email{ paillard@uark.edu}

\title{Photogalvanic shift currents in BiFeO\textsubscript{3}-LaFeO\textsubscript{3} superlattices}


\begin{document}

\begin{abstract}
Designing materials with controlled photovoltaic response may lead to improved solar cells or photosensors. In this regard, ferroelectric superlattices have emerged as a rich platform to engineer functional properties. In addition, ferroelectrics are naturally endowed with a bulk photovoltaic response stemming from non-thermalized photo-excited carriers, which can overcome the fundamental limits of current solar cells. Yet, their photovoltaic output has been limited by poor optical absorption and poor charge collection or photo-excited carrier mean free path.
We use Density Functional Theory and Wannierization to compute the so-called Bulk Photovoltaic shift current and the optical properties of BiFeO\textsubscript{3}/LaFeO\textsubscript{3} superlattices. We show that, by stacking these two materials, not only the optical absorption is improved at larger wavelengths (due to LaFeO\textsubscript{3} smaller bandgap), but the photovolgavanic shift current is also enhanced compared to that of pure BiFeO\textsubscript{3}, by suppressing the destructive interferences occurring between different wavelengths.
\end{abstract}

\maketitle

\section{Introduction}
\large

The bulk photovoltaic (or more accurately, photogalvanic) effect (BPVE)~\cite{Sturman1992} has harbored the promise of breaking the fundamental thermodynamic limit which impedes the yield of current junction-based solar cells, known as the Shockley-Queisser limit~\cite{Shockley1961}. One of the first requirement for the BPVE is the absence of inversion symmetry. Therefore, ferroelectrics, which show a spontaneous electrical polarization that can be switched by electric fields, have been deemed particularly promising photovoltaic materials as they naturally break inversion symmetry. In fact, Spanier \textit{et al.} showed that the Shockley-Queisser barrier could be overcome in a thin film of barium titanate~\cite{Spanier2016}, a prototypical ferroelectric materials. However, it has proven difficult to reach power conversion efficiencies beyond a few percents with ferroelectric-only based solar cells. This is due, in part, to the large bandgap of most ferroelectrics which start absorbing in the ultraviolet part of the solar spectrum, as well as the usually low carrier mobility exhibited by conventional ferroelectrics~\cite{Paillard2016,Frye2023}.

To circumvent these problems, many have attempted to partially modify the chemical structure of the active absorbing element. 
For instance, by making a solid solution involving ferroelectric KNbO\textsubscript{3} and oxygen deficient BaNi$_{1/2}$Nb$_{1/2}$O$_{3-\delta}$, the bandgap of KNbO\textsubscript{3} is strongly reduced down to 1.4 eV~\cite{Grinberg2013}.
Alternatively, co-doping methods, used to improve optical absorption while limiting leakage and maintaining ferroelectric properties, have been explored but often result in localized intragap states with very limited dispersion \cite{Hao2022}. 
Perhaps most promisingly, Nechache \textit{et al.} reached a 8.1\% power conversion efficiency by stacking 3 layers of Bi(Fe,Cr)O\textsubscript{3} thin films with varying Cr content to enhance light absorption~\cite{Nechache2015}.
This latter study, combined with the improved photovoltaic efficiency of barium titanate thin films~\cite{Zenkevich2014} and demonstrated enhanced collection of photo-induced charges by Atomic Force Microscopy tips~\cite{Spanier2016,Alexe2011}, strongly indicate that superlattice architectures, which are periodical repeated stackings of nanolayers, may help (1) enhancing optical absorption by combining materials with different bandgaps and (2) increase the collection of photocurrents created by the bulk photovoltaic effect.

In this work, we investigate the BPVE in superlattices (SLs) of BiFeO$_3$ (BFO) and LaFeO$_3$ (LFO) by means of Density Functional Theory (DFT) calculations. The choice of these two perovskites is dictated by several natural reasons. Firstly, BFO is one of the few ferroelectrics which absorb visible light with its typical bandgap of 2.7~eV, showing for instance good piezo-photocatalytic~\cite{Amdouni2023} or photovoltaic responses~\cite{Choi2009,Yang2009}, as well as a predicted high shift current response compared to other classical ferroelectrics~\cite{Rappe_BFO_2012}. Secondly, LFO has a smaller bandgap of about 2.34~eV~\cite{Scafetta_2014}, has been grown in superlattices with BFO~\cite{Zanolli_2014, Carcan2018,Gu_2024} exhibiting ferroelectric features, in particular at short periods~\cite{Zanolli_2013}. We thus expect that BFO/LFO SLs should exhibit some degree of improved visible absorption and BPVE thanks to the complementarity of the BFO and LFO nanoscale layers.

\section{Systems and methods}

The linear BPVE contains two main contributions~\cite{Sturman1992}: a ballistic current, caused by asymmetry in the distribution of photo-induced carriers in momentum space, and a shift current \cite{Resta2024} caused by asymmetry in the spatial location of photo-induced carriers after photo-excitation. Young \textit{et al.}~\cite{Rappe_BFO_2012,Rappe_BTO_2012} have developed methods to  calculate the latter from DFT. A third order tensor, $\sigma_{ijk}$,  describes the relationship between the shift photogalvanic current density in direction $i$, $j_i$, the intensity $I$ and polarization vector $\bm{e}$ of a linearly polarized electromagnetic wave of pulsation $\omega$,
\begin{equation}
    j_i = \sigma_{ijk} I e_j e_k 
    \label{eq:bpve}
\end{equation}
%
%
sTo calculate this tensor in BFO/LFO SLs, we performed DFT spin-polarized calculations as implemented in Quantum Espresso \cite{Giannozzi_2009} in the projected augmented wave (PAW) scheme~\cite{Blochl_1994} at the PBEsol \cite{Perdew_2008} level of approximation. We included the rotationally invariant U correction, in the simplified Dudarev approach, with $U = 4.5$~eV~\cite{Zanolli_2014} to represent the on-site Coulomb interaction on Fe-ions belonging to the BFO layer, and $U=3.65$~eV on the Fe-ions belonging to the LFO layer. This choice allows to maintain a similar bandgap difference between bulk BFO (which we calculate at 2.26~eV here) and bulk LFO (1.83~eV calculated bandgap).
We relaxed pure BFO and LFO until the maximum force was smaller than 10$^{-3}$ eV/\AA, sampling the Brillouin zone with a Monkhorst-Pack mesh with a density of at least 110 kpoints$\times\AA^3$. 
With this choice of U, the relaxed ground state for BFO and LFO are described respectively by a pseudo-cubic lattice parameter $a=3.977$ \AA\, and an angle of $\alpha = 89.65$~degree and by an orthorhombic Pbnm phase whose primitive vectors are $a=5.531$ \AA,  $b=5.591$ \AA, $c=7.838$ \AA, in good agreement with the literature \cite{Zanolli_2014}. 

We then considered the superlattices as if they were epitaxially grown on (001)-SrTiO$_3$, which is a common substrate for growing BFO films and BFO/REFO superstructures \cite{Khaled_2021,Maran_2014,Maran_2016}.
We create the SL by alternating layers of BFO and LFO along the c direction of LFO.
In this orientation, the in-plane pseudo-cubic vectors of BFO form a 45 degree angle with the in-plane primitive vectors of the superlattice, and the polarization is now directed along the $[101]$ direction, as reported in Figure \ref{fig:structure}. The superlattice primitive vectors in Figure \ref{fig:structure} are chosen as the lab reference frame, and all subsequent tensor components (optical dielectric permittivity,shift current tensor) refer to these axes.
The mechanical effect of the substrate was simulated by fixing the in-plane lattice constant of the multi-layer to the one of cubic SrTiO\textsubscript{3} (previously relaxed with DFT, lattice parameter of 3.891~\AA in its cubic unit cell), allowing all the remaining degrees of freedom to relax. 
In this configuration the BFO in-plane pseudo-cubic lattice parameters are strained by -2\% when comparing to the BFO relaxed bulk structure, while the \textbf{a} and \textbf{b} lattice parameters of LFO are strained by -0.6\% and -1.6\% respectively.
%
It is known that STO substrates inhibit the cycloidal order in BFO films \cite{sando_2013} leading rather to G-type antiferromagnetic (AFM) order.
We thus performed spin-polarized simulations without spin-orbit coupling (SOC) to correctly describe the antiferromagnetic G-type magnetic ordering of the two pure perovskites. Past theoretical studies have shown that good agreement of the shift current with experiment could be obtained without SOC~\cite{Young2012b}.
After relaxation, the self-consistent charge densities were employed as input to obtain Kohn-Sham wave functions from non-self consistent calculation on denser k-point meshes (of the order of $10^3$ kpoints$\times\AA^3$).
The resulting Kohn-Sham wave functions were then processed through Wannier90 \cite{Pizzi2020} to obtained the corresponding set of maximally-localized Wannier orbitals. A mixture of chemically sound atomic orbitals (O {\it p}, Fe {\it p} and {\it d}, Bi {\it p}, La {\it p}) were employed as an initial guess to derive the Wannier orbitals. An example of the bandstructure reconstructed from Wannier orbitals is shown in Figure \ref{fig:bandstructure_optics}. 
We then obtained the electrons transport properties by post-processing the Wannier orbitals using the method described in Ref.~\cite{Azpiroz2018} and extracted the shift-current contribution to the BPVE in these SLs, as well as the Wannier-orbital based Kubo-formulation to obtain optical conductivity and dielectric response.
We focus, in the following, on BFO\textsubscript{1}/LFO\textsubscript{1} SLs structures.

\section{Results}

\subsection{Superlattice structure}
\begin{figure}
\centering
\includegraphics[scale=1]{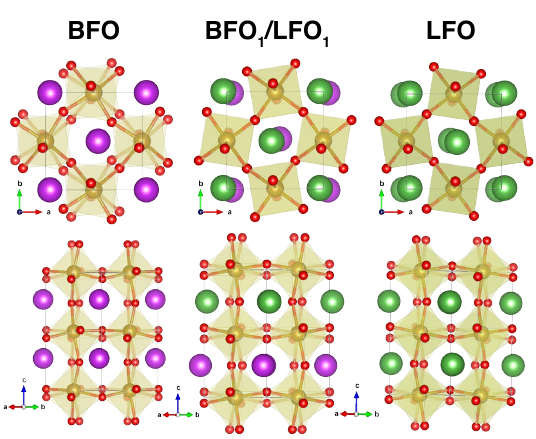}
\caption{Calculated cells for strained BFO (left), BFO\textsubscript{1}/LFO\textsubscript{1} (middle) and LFO (right); oxygen atoms are depicted in red, iron atoms in gold, bismuth in purple, and lanthanum atoms are depicted in green.}
\label{fig:structure}
\end{figure}

Let us first describe the relaxed structure of the different components of the BFO\textsubscript{1}/LFO\textsubscript{1} SLs, compared to pure BiFeO\textsubscript{3} and LaFeO\textsubscript{3}. As can be observed in Figure~\ref{fig:structure}, BFO bi-axially strained at the STO lattice constant exhibits its well-known R-phase structure~\cite{sando_2013}, characterized by a total polarization lying along the diagonal of the perovskite cube (which corresponds, in our chosen reference frame, to the $[101/2]$ direction) estimated to be 81.9 $\rm \mu C/cm^2$ and an oxygen octahedra tilt pattern $a^-a^-a^-$ in Glazer notation~\cite{Glazer_1972}. 
On the other hand, strained LFO (see right panel in Figure~\ref{fig:structure}) retains its $Pnma$ structure, characterized by antipolar displacements along $[100]$ direction and a tilt pattern of the form $a^-a^-c^+$. When relaxing the BFO\textsubscript{1}/LFO\textsubscript{1} superlattice, the most stable phase shows $a^-a^-c^+$ tilt patterns, in agreement with previous reports~\cite{Zanolli_2013,Xu_2024}, with no out-of-plane polarization but in-plane polarization along the $[100]$ axis due to the difference in the amplitude of cationic motion of La and Bi ions (see Figure~\ref{fig:structure}). The space group of the relaxed superlattice is $Pc$.

\subsection{Electronic bandstructure and optical properties}

\begin{figure*}
\begin{center}
 \includegraphics[scale=0.85]{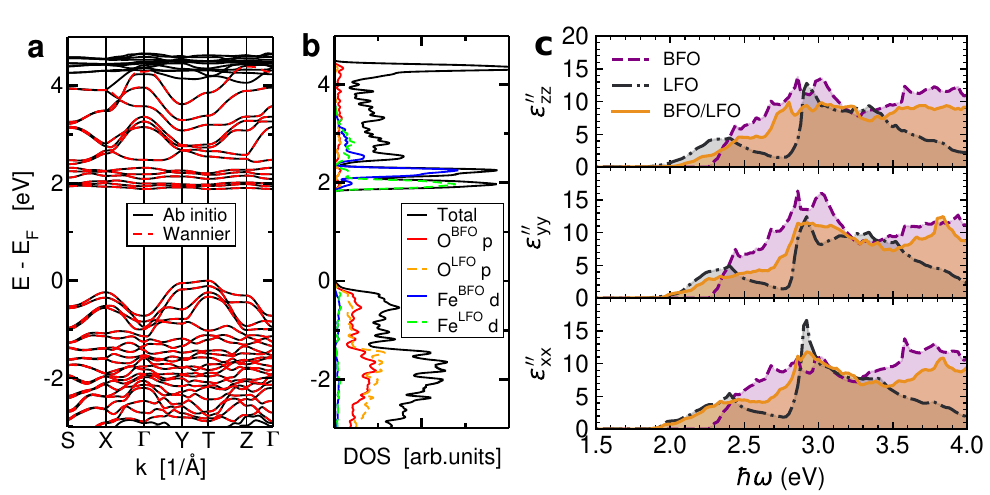}
 \caption{(a) Band structure of BFO\textsubscript{1}/LFO\textsubscript{1} SL calculated from DFT (black solid lines) and from Wannier interpolation (red dashed lines). (b) Layer-by-layer decomposed Density of States in BFO\textsubscript{1}/LFO\textsubscript{1} SL: dashed lines corresponds to orbitals in the LFO layer, continuous lines to those in BFO layer. (c) imaginary part of the longitudinal dielectric relative permittivity shows that the onset of BFO/LFO (yellow plain line) absorption occurs, as in LFO (black dashed-dotted lines), at lower photon energy compared to bulk BFO (purple dashed lines).
 }
 \label{fig:bandstructure_optics}
\end{center}
\end{figure*}

Next, we focus on the electronic bandstructure and ensuing optical properties, calculated in the Independent Particle Approximation (see Methods and Systems section). In Figure~\ref{fig:bandstructure_optics}a, we plot the bandstructure of BFO\textsubscript{1}/LFO\textsubscript{1}. It shows that the electronic bandgap of 1.88 eV  is indirect between the $T$ and $\Gamma$ point, although an almost equally large direct bandgap is present at the $T$ point. Inspection of the atomic-decomposed Density of States (DOS) indicates that the bottom of the conduction band at $\Gamma$ mainly consists of Fe $d$ states belonging to the FeO\textsubscript{2} plane closer to the LaO plane (i.e. to the LFO layer). In contrast, the top of the valence band, located at the $T$ point in the Brillouin zone, is primarily made of O $2p$ states belonging to the FeO\textsubscript{2} planes. 
Pure compressed BFO and LFO have indirect band gap too, as reported in Supplementary Information and consistent with the litterature~\cite{Paillard2016b,Scafetta_2014}. The conduction (valence) bands minimum (maximum) sit at $\Gamma$ (along the S-X line) in BFO, and at $\Gamma$ (X) in LFO.
The orbital character is preserved when the SL is formed.
%

Figure~\ref{fig:bandstructure_optics}c compares the optical dielectric permittivity of BFO\textsubscript{1}/LFO\textsubscript{1} SL, strained LFO and BFO bulk. Strained LFO, as expected, starts significantly absorbing at lower photon energies (near about 1.9~eV) compared to bulk BFO (which starts absorbing near 2.25~eV). Interestingly, the optical absorption of BFO\textsubscript{1}/LFO\textsubscript{1} SL is dominated by the LFO layer at low photon energies (from 1.9~eV to 2.25~eV). Therefore, the SL architecture successfully manages to improve optical absorption at lower photon energies compared to pure BFO. In contrast, at larger photon energies ($>2.3$~eV), the imaginary part of the relative dielectric permittivity is significantly degraded compared to pure BFO, indicating that optical absorption will be poorer than BFO at large photon energies. Yet, the presence of BFO improves optical absorption over LFO in BFO/LFO SLs.

\subsection{Shift current photogalvanic tensor}

\begin{figure*}
\begin{center}
\includegraphics[scale=1]{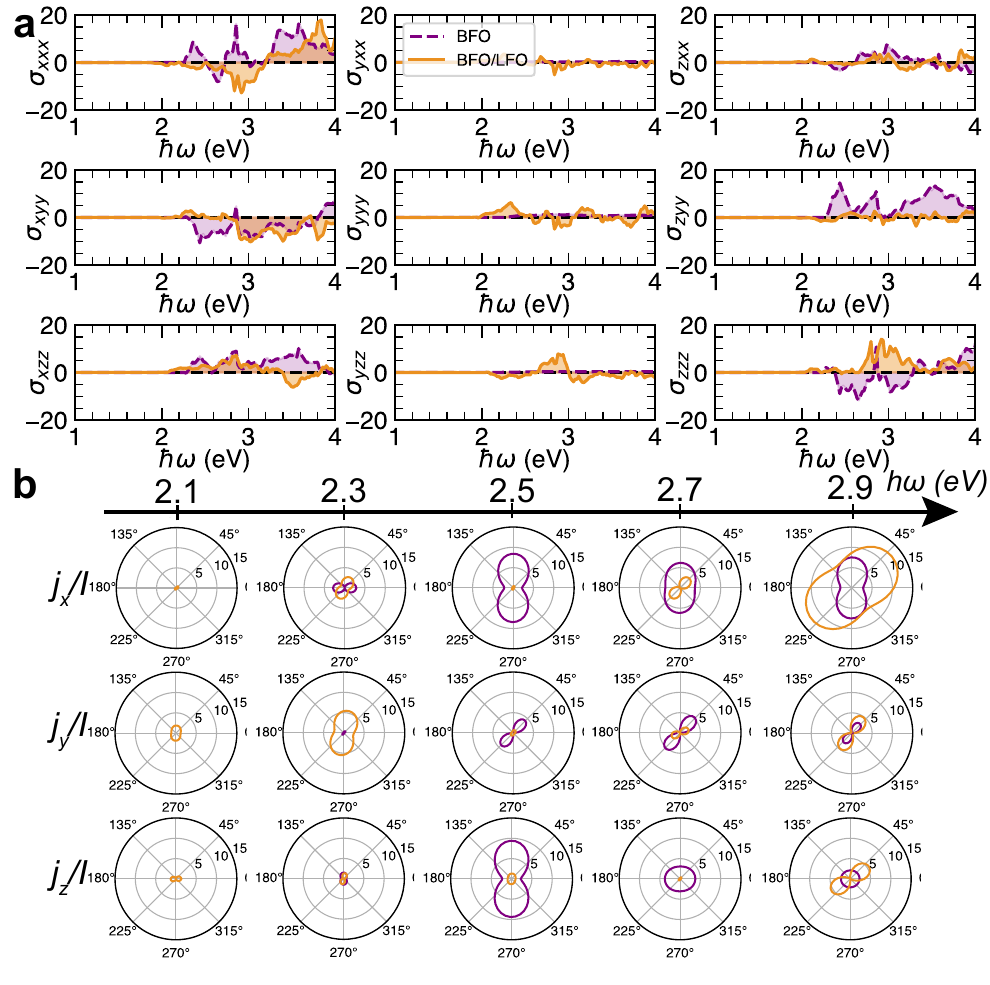}
\caption{(a) Shift current tensor components (in $\rm \mu A \cdot V\textsuperscript{-2}$) of BFO\textsubscript{1}/LFO\textsubscript{1} (yellow) and bulk BFO (purple) corresponding to light linearly polarized along the lab reference frame axes. (b) polar plot showing the magnitude of the current density response $j_i$ to light polarized in the (xy) lab reference frame for BFO\textsubscript{1}/LFO\textsubscript{1} (yellow) and bulk BFO (purple).}
\label{fig:sc_tensor}
\end{center}
\end{figure*}

We now focus on the BPVE response of our BFO/LFO SLs. In Figure~\ref{fig:sc_tensor}, we plot the shift current tensor $\sigma_{ijk}$ (which relates the photogalvanic current density to the incident light intensity and polarization, see Eq.~\ref{eq:bpve}) for BFO\textsubscript{1}/LFO\textsubscript{1} SL and bulk BFO (LFO is centrosymmetric, and thus does not exhibit a BPVE response). Interestingly, as already observed in the dielectric permittivity in Figure~\ref{fig:bandstructure_optics}c, the shift current photo-response in BFO\textsubscript{1}/LFO\textsubscript{1} SL starts at lower photon energies. For instance, in Figure~\ref{fig:sc_tensor}a, the $\sigma_{yyy}$ component of the shift current tensor has a finite value (near 2-6~$\mu$A/V\textsuperscript{2}), which is 20 times as large as the response of BFO for photon energies in the range 2.1-2.3~eV, \textit{i.e.} right before the absorption edge of pure BFO.

In contrast, for photon energy larger than the calculated bandgap of BFO ($\approx 2.26$~eV), the magnitude of the BPVE shift current tensor in BFO\textsubscript{1}/LFO\textsubscript{1} is on par or slightly smaller than BFO. To further probe the photovoltaic response of BFO\textsubscript{1}/LFO\textsubscript{1}, we depict polar plots in Figure~\ref{fig:sc_tensor}b representing the absolute value of the photogalvanic shift current density response at selected photon energies. We assume that the visible light is polarized in the (xy) plane, i.e. that the polarization vector of the light is of the form $\bm{e} = (\cos\phi,\sin\phi,0)$, which essentially corresponds to the case of a through-thickness device. The results in Figure~\ref{fig:sc_tensor}b clearly show that up to 2.3~eV, the shift current response of the SL is superior to that of pure BFO. This is no longer the case for photon energies of 2.5 or 2.7~eV. At further photon energies, such as 2.9~eV, a strong peak in the shift current components $\sigma_{xxx}$, $\sigma_{xyy}$ and $\sigma_{zzz}$ of the SL yields a large photocurrent response compared to BFO, where destructive interferences strongly reduce the photovoltaic tensor component magnitude. Note that the shift current response is shifted by about 0.7~eV when realizing calculations with the HSE hybrid functional~\cite{Heyd2003} (see Supplementary Information).

\begin{figure}
\begin{center}
\includegraphics[scale=1]{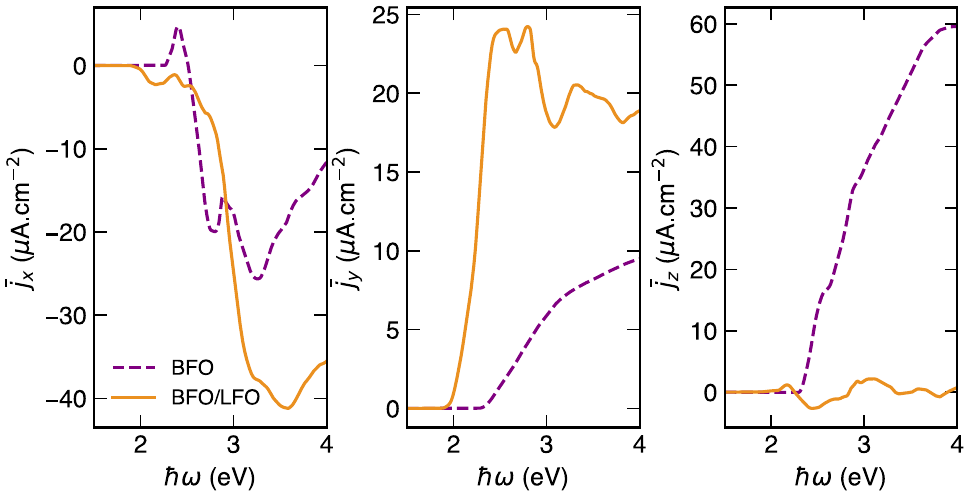}
\caption{Integrated shift photogalvanic current using the Planck distribution as a weighting function in BFO\textsubscript{1}/LFO\textsubscript{1} (yellow) and BFO (dashed purple), assuming unpolarized light.}
\label{fig:weighted_average}
\end{center}
\end{figure}

Interestingly, various components of the BPVE shift current tensor no longer alternate sign for photon energies in the range 2-3~eV (see for instance $\sigma_{xxx}$ or $\sigma_{zzz}$ in Figure~\ref{fig:sc_tensor}a). This is particularly interesting since, in the case of BFO, several components such as $\sigma_{xxx}$ change sign with different photon energies, resulting in destructive interferences which limit the BPVE output. To further quantify the potential advantage of BFO/LFO SLs, we compute the weighted average of each components of the BPVE, with the Planck distribution as a weighted average. Specifically, we compute

\begin{equation}
\bar{\sigma}_{ijk}(\hbar\omega) = \frac{2}{c\varepsilon_0}\int_0^{\hbar\omega} dE B(E,T)  \sigma_{ijk}(E),
\label{eq:weighted_average}
\end{equation}

where $\varepsilon_0$ is the vacuum permittivity, $c$ the speed of light, $B(E,T) =  \frac{2}{h^3c^2} \frac{E^3}{e^{\frac{E}{k_B T}} - 1}$ is the black body spectral radiance in W.m\textsuperscript{-2}.sr\textsuperscript{-1}.J\textsuperscript{-1}, 
$h$ the Planck constant, $k_B$ the Boltzmann constant and $T\approx 5,777$~K the temperature of the sun, approximated as a black body. Essentially, $\bar{\sigma}_{ijk}$ acts as a sort of Figure of Merit which integrates the shift current tensor to appropriately sum positive or destructive interferences in the photogalvanic response at different photon energies.
Note that technically, we apply a small scissor shift correction to $\sigma_{ijk}(E)$ when computing $\bar{\sigma}_{ijk}(\hbar\omega)$, as we know that the calculated bandgaps of BFO and LFO are underestimated by approximately 0.5~eV compared to experimental values. The quantity $\bar{\sigma}_{ijk}(\hbar\omega)$, in the limit $\hbar\omega \rightarrow \infty$, approximates the integrated output photogalvanic current density (see Supplementary Information), and thus integrates the destructive/constructive interferences occurring due to the summation over multiple wavelengths. 
To further simplify the matter, we depict, in Figure~\ref{fig:weighted_average}, the quantities $\bar{j}_i(\hbar\omega) = \frac{1}{2} \left(\bar{\sigma}_{ixx}(\hbar\omega) + \bar{\sigma}_{iyy}(\hbar\omega)\right)$, which approximate the photogalvanic current density under an unpolarized light traveling in the $(001)$ direction of the SL. Figure~\ref{fig:weighted_average} shows an important feature: the photogalvanic outputs $\bar{j}_{x}$ and $\bar{j}_{y}$ (i.e. for in-plane current densities generated by unpolarized light) are almost twice as large as those in pure BFO.
This is due, in part, to the fact that the shift current tensor components $\sigma_{xxx}$ and $\sigma_{yyy}$ BFO\textsubscript{1}/LFO\textsubscript{1} do not change sign over a significant range of photon energy where the Planck distribution. One may wonder whether this is related to the smaller in-plane polarization in BFO\textsubscript{1}/LFO\textsubscript{1} (about 10 $\rm \mu C/cm^2$, calculated by the Berry phase method~\cite{King-Smith_1993}) compared to bulk BFO (whose polarization has been measured to exceed 100~$\rm \mu C/cm^2$~\cite{Lebeugle2007}). 
Although there is no straightforward relationship between the electrical polarization and the shift current tensor~\cite{Wang_2016, Young2012}, previous theoretical works have shown how some superlattice-like structures reduce the polar ionic cations and may lead to a strong enhancement of the integrated BPV shift current output~\cite{Wang2016_nanolayers}. 
Thus, the smaller polarization and anisotropic distortion of our BFO\textsubscript{1}/LFO\textsubscript{1} SL likely favor  constructive interferences compared to BFO. In contrast, the current generated in the out-of-plane direction of the SL is much smaller than in pure BFO. As our SL features no out-of-plane polarization component in the $z$ direction, we essentially  recover a mirror symmetry along $z$ which would prevent the emergence of a photogalvanic current when light polarized along an in-plane direction is absorbed, as previously noted in a different nanolayered ferroelectric oxide in Ref.~\cite{Wang2016_nanolayers}. 
Note also that the giant bulk photovoltaic effect found in the prototypical ferroelectric BaTiO\textsubscript{3} reported photocurrent density smaller than 10~$\mu$A.cm\textsuperscript{-2}~\cite{Zenkevich2014}. In contrast, BFO and LFO are both known to have smaller band gaps, and our calculated shift current tensor components such as $\sigma_{xxx}$ in Figure~\ref{fig:sc_tensor} reach larger magnitude at lower photon energy compared to previous theoretical works~\cite{Young2012} if one accounts for the difference in bandgaps. We thus anticipate that, given the shape of the black-body weighting function in Equation~\ref{eq:weighted_average}, BFO\textsubscript{1}/LFO\textsubscript{1} would yield superior photovoltaic current density output compared to BaTiO\textsubscript{3} thin films.

Finally, one may wonder whether other BFO\textsubscript{n}/LFO\textsubscript{m} may yield larger photocurrent density than BFO\textsubscript{1}/LFO\textsubscript{1}. We argue that this is unlikely. Indeed, previous works indicate that for $m\geq 2$, computationally tractable BFO\textsubscript{n}/LFO\textsubscript{m} superlattices are centrosymmetric~\cite{Zanolli_2014}, thus preventing the shift current effect due to the presence of inversion symmetry. On the other hand, one may want to explore BFO\textsubscript{n}/LFO\textsubscript{1} superlattices. Yet, it can be expected that, for large $n$, the photocurrent density will not be significantly improved due to the lower proportion of LFO, and thus poorer absorption at large wavelengths.

\section{Conclusion}

Using first-principle calculations, we show how stacking nanolayers of a ferroelectric, BiFeO\textsubscript{3}, with a dielectric absorbing at lower photon energy (LaFeO\textsubscript{3}) not only increases optical absorption but may also  improve the Bulk Photovoltaic shift current output. The latter improvement results from the subtle reduction in cationic displacements which, as in Ref.~\cite{Wang_2016}, leads to the emergence of a larger range of photon energies where the shift current maintains the same direction. Our work further highlights how ferroelectric nanostructures, such as superlattices, may be used to engineer and improve the photovoltaic response of ferroelectrics. 
We hope that the current theoretical work will stimulate experimentalists to explore the shift current photovoltaic response of ferroelectric superlattices. The material-based, nanostructuring strategy adopted in this work is a first step toward evaluating the potential of ferroelectric superlattices for photogalvanic conversion. Future works, including the effect of bi-axial strain, temperature, as well as a system approach involving the nature of electrical contacts are needed to further optimize the photovoltaic output of multiferroic-based superlattices.

\section{Supporting information}
The supporting information contains: the absorption coefficient and the Glass coefficient~\cite{Ibanez2020}; the electrons bands and the atomic-orbital projected DOS of pure perovskites; the study of the electronic band gap computed with Hybrid functional; the comparison of the full shift-conductivity tensors of pure BiFeO$_3$ and  BFO\textsubscript{1}/LFO\textsubscript{1} for different percentages of exact exchange and Hubbard U corrections applied.

\section*{Acknowledgement}
This work was supported by Agence Nationale de la Recherche under grant agreement no. ANR-21-CE24-0032 (SUPERSPIN). This work was performed using computational resources from GENCI- TGCC (grant 2024-A0150912877) as well as from the “Mésocentre” computing center of Université Paris-Saclay, CentraleSupélec and École Normale Supérieure Paris-Saclay supported by CNRS and Région Île-de-France (https://mesocentre.universite-paris-saclay.fr/). 

\bibliography{biblio}

\end{document}